\documentclass{optica-article}

\journal{opticajournal} 

\articletype{Research Article}

\usepackage{lineno}

\usepackage{graphicx}
\usepackage{dcolumn}
\usepackage{bm}
\usepackage{subfig}
\usepackage{braket}
\usepackage{xcolor}
\usepackage{ulem}
\usepackage{float}
\usepackage{xcolor}

\begin{document}

\title{Hong-Ou-Mandel effect with two frequency-entangled photons of vastly different color}

\author{Felix Mann$^{1,*}$, Helen M. Chrzanowski$^{1}$, Marlon Placke$^{1}$, Felipe Gewers$^{1}$, Sven Ramelow$^{1,2}$}

\address{\authormark{1}Institut f\"ur Physik, Humboldt-Universit\"at zu Berlin, Newtonstr. 15, 12489 Berlin, Germany\\
\authormark{2}Ferdinand-Braun-Institut (FBH), Gustav-Kirchhoff-Straße 4, 12489 Berlin, Germany}

\email{\authormark{*}felixmann@physik.hu-berlin.de} 

\begin{abstract*}
In the original formulation of the Hong-Ou-Mandel (HOM) experiment, when two otherwise indistinguishable photons are incident upon the two input ports of a balanced beam splitter, they coalesce, always leaving via the same output port. It is often interpreted that this interference arises due to the indistinguishability of the single photons at the beam splitter; the situation, however, is often more nuanced. Here, we demonstrate an analog of HOM interference between two photons of completely different color. To do so, we utilize a quantum frequency converter based on sum- and difference-frequency generation as an `active' beam splitter -- coupling frequency-entangled red and telecom single photons with an octave-spanning energy difference of 282 THz. We achieve an uncorrected two-photon interference visibility beyond 90\%. This work presents the first demonstration of HOM interference between two single photons of distinctly different color, deepening our understanding of what underlies quantum interference. It also suggests a novel approach to interfacing photonic qubits in heterogeneous quantum systems where frequency conversion and quantum interference are unified.
\end{abstract*}

\section{Introduction}
The Hong-Ou-Mandel (HOM) effect~\cite{hom} is a well-known quantum phenomenon where -- in its canonical formulation -- two otherwise indistinguishable photons incident upon the two input ports of a balanced lossless beam splitter will always leave one of the two output ports together. This bunching effect is enabled by destructive interference of the probability amplitudes associated with the two photons leaving via different ports.

Beyond its historical importance in underpinning a necessarily quantum mechanical description of light, the HOM effect is central to tasks in photonic quantum information processing~\cite{computing,multiphoton}. Two-photon interference, when combined with the probabilistic nonlinearity of single photon detection, allows -- at least in principle -- for the construction of a universal quantum computer with single photons and linear optics~\cite{KLM}. It is also the elemental building block of the complex, multi-photon interference that underlies classically intractable tasks such as BosonSampling~\cite{AaronsonBS}. It facilitates the Bell-state measurements necessary for state teleportation and entanglement swapping~\cite{multiphoton} that interlink nodes of a quantum network~\cite{kimble,hanson}. And, in sensing with quantum light, it enables ultra-precise timing measurements~\cite{attosecond}, dispersion-free optical coherence tomography (OCT)~\cite{OCTLarchuk} or the generation of N-photon entangled states of light that offer the promise of interferometric phase-sensing beyond the shot-noise-limit~\cite{KokNOON}.

Due to its sensitivity to the indistinguishability of the two input photons, HOM interference is nowadays near-universally employed as a test of photon indistinguishability~\cite{santori}. Here, indistinguishability refers to a property of the two incoming photons, namely that no measurement could -- even in principle -- distinguish between them. Owing in part to the ubiquity of HOM interference as the benchmark for the quality of single photon sources, the indistinguishability of two photons is often considered synonymous with their ability to interfere. This intuition, however, neglects the role of measurement and is especially found wanting when considering interference of more than two photons. In the past decade, theoretical \cite{dittel2021} and experimental~\cite{tillmann2015,Menssen2017,agne2017,jones2020} efforts have considered multi-photon generalizations of HOM interference, with the emerging consensus that this indistinguishability of photons is not synonymous with bunching~\cite{Seron2023}.

In generality, the necessary condition to observe quantum interference in a specific experiment is whether an outcome can occur in several non-exclusive alternative ways~\cite{feynmanlectures,mandel_coherence}. Quantum interference is therefore best understood as the indistinguishability of alternative paths in quantum state space and not the indistinguishability of the photons themselves. In the original HOM experiment, the output state associated with a coincidence count can arise from either both photons being reflected or both being transmitted at the beam splitter~\cite{multiphoton}. Without the ability to label the photons in their internal degrees of freedom, these two alternative paths are indistinguishable and destructively interfere. Subsequent experiments elucidated the necessity of the path interpretation, showing that the photons in a HOM experiment need not meet at the beam splitter at the same time~\cite{pittman} nor must they share the same polarization~\cite{kwiat}, provided the information associated with which paths the photons took is erased in detection.

\begin{figure}[H]
\centering
\includegraphics[width=8.6cm]{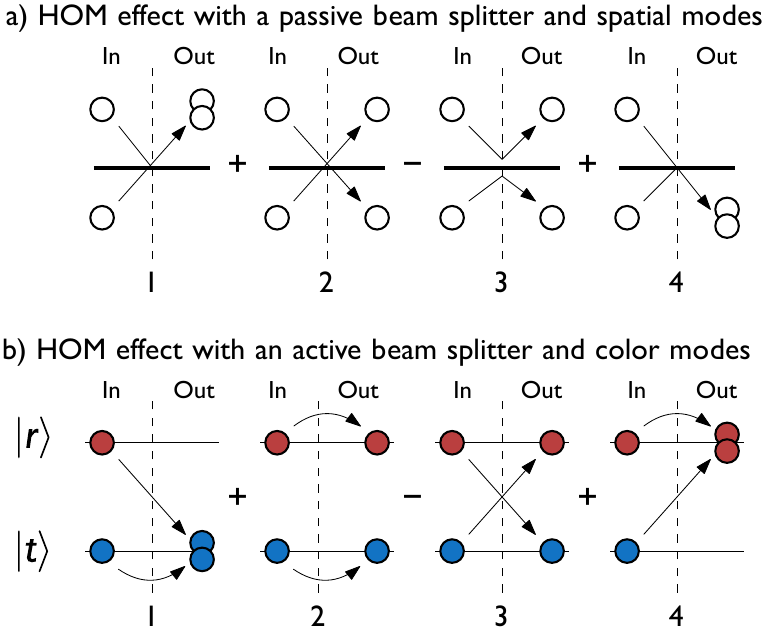}
\caption{The principle of the Hong-Ou-Mandel (HOM) effect with a) photons of the same color and with b) photons of different color. In both scenarios, the probability amplitudes for four possible paths need to be considered when calculating the state at the output. Paths 2 and 3 are indistinguishable and interfere destructively and consequently -- in the case of an ideal implementation and a balanced beam splitter -- the measurement outcomes associated with the output photons having different color disappear.}
\label{fig:scheme}
\end{figure}

In the idealized original HOM experiment, a passive symmetric beam splitter couples the spatial modes of two otherwise indistinguishable photons. Here, we consider the two-color analogue, with two photons, distinguishable only in their very different wavelengths, coupled via an 'active' beam splitter. In the original theory proposal of Raymer et al., the active beam splitter was envisaged as a moving mirror, imparting a frequency shift to the input modes via a relativistic Doppler effect~\cite{raymer}; though the authors also outlined more practicable implementations via an acousto-optic modulator (AOM) (equally an electro-optic modulator (EOM)) or a quantum frequency converter~\cite{kumar}. The latter, which can be implemented via three-wave-mixing (TWM) or four-wave-mixing (FWM), also allows for a practical implementation across very different wavelengths, coupling modes of different color with a transition probability given by the conversion efficiency. Fig.~\ref{fig:scheme} illustrates HOM interference with a passive or an active beam splitter. The quantum mechanical description for the passive and active beam splitters are mathematically equivalent~\cite{raymer,review_gaeta,campos}. See Supplement 1 for further details on this. 

Variations of this effect have been demonstrated in several different experimental settings: using difference -- and sum -- frequency generation (DFG/SFG) to couple a single photon and a weak laser pulse separated by 187 THz \cite{frequencyHOM_pulse}, using FWM to couple photon pairs separated by 805 GHz~\cite{frequencyHOM_sfwm}, and utilizing a fast EOM to couple photon pairs separated by 100 GHz~\cite{frequencyHOM_twosources} and 22 GHz~\cite{frequencyHOM_linear} respectively. A demonstration in the microwave regime has been realized with a separation of 6 GHz~\cite{nguyen2012}. Further, this effect has also been extended to Bell state measurements, albeit for comparatively small photon energy differences~\cite{colorBSM}.

Here, we demonstrate HOM interference between two photons of vastly different color using a low-noise quantum frequency converter \cite{mann2} based on  TWM - more specific SFG/DFG - as an active beam splitter. The two photons -- one red (637 nm or 471 THz) and the other at telecommunication wavelengths (1587 nm or 189 THz) -- originate from a photon pair source based on spontaneous parametric down-conversion (SPDC) and have an octave-spanning energy difference of 282 THz, corresponding to the energy of the 1064 nm pump photon driving the conversion. This energy difference is about 350 times larger than that of photon pairs in previous experiments~\cite{frequencyHOM_linear,frequencyHOM_sfwm,frequencyHOM_twosources,review_gaeta}.\\
Alongside emphasizing the fundamental principles that underlie quantum interference, this approach allows one to combine frequency conversion with quantum interference for heterogeneous quantum systems in a single process.
\section{Experimental setup}
\begin{figure}[H]
\centering
\includegraphics[width=8cm]{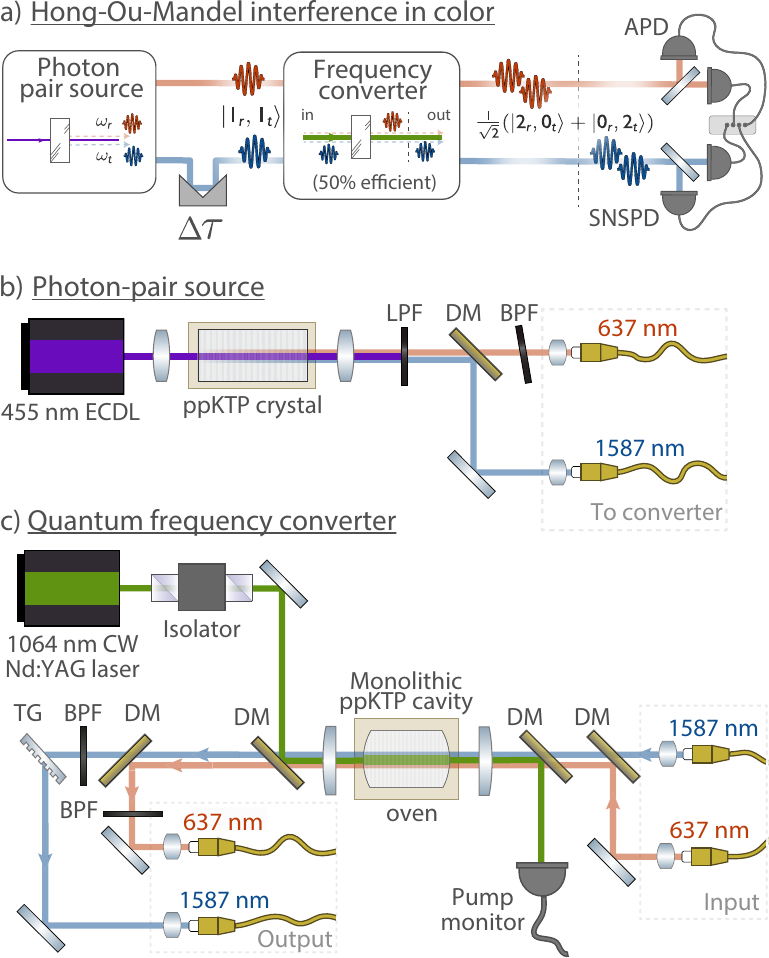}
\caption{Experimental setup for the Hong-Ou Mandel effect in color. A blue pump photon decays via spontaneous parametric down-conversion into a photon pair at red and telecommunication wavelengths, with the photon frequencies matched to the transition frequencies of a quantum frequency converter. A tunable relative time delay $\Delta\tau$ is introduced for the photon pair in the telecom arm. Subsequently, both photons are aligned into a quantum frequency converter operating at 50\% conversion efficiency. To resolve the coincidence events within both colors, the red and telecom output modes are further split via ordinary balanced beam splitters before detection. Coincidence counts are recorded with four single photon detectors as a function of the relative time delay $\Delta\tau$.}
\label{fig:setup}
\end{figure}

An overview of the entire experimental setup is provided in Fig.~\ref{fig:setup}. The photon pair source (Fig.~\ref{fig:setup}b) employed a periodically-poled Potassium titanyl phosphate (ppKTP) crystal pumped with a continuous-wave (CW) external-cavity diode laser at 455 nm to generate photon pairs at 637 and 1587 nm via a quasi-phase-matched type-0 SPDC process (poling period of 6.207 $\mu\mbox{m}$). The ppKTP crystal was 31 mm long with a resulting bandwidth of $\Delta\nu_{s} \approx 75\mbox{ GHz}$ (FWHM), well-matched to the conversion bandwidth of the frequency converter. The `active' beam splitter was realized via a quantum frequency converter (Fig.~\ref{fig:setup}c) based on SFG/DFG in a type-0 ppKTP crystal, interconnecting the wavelengths of 637 and 1587 nm via a bright 1064 pump~\cite{mann2}. The 20 mm long ppKTP crystal has a poling period of 15.75 $\mu\mbox{m}$, resulting in a conversion bandwidth of $\Delta\nu_{c} \approx 110\mbox{ GHz}$ (FWHM). To achieve sufficiently high conversion efficiencies, the 3 W Nd:YAG CW pump laser was resonantly enhanced via a monolithic cavity formed by polished and coated end facets of the conversion crystal, yielding a power enhancement factor of about 50. With the crystal temperature stabilized below 1 mK, an intrinsic thermal feedback passively locks the system on resonance without the need for any additional active feedback stabilization. Owing to the high poling quality that is characteristic of bulk ppKTP~\cite{mann1}, the converter demonstrates comparatively low pump-induced noise at the interconnected wavelengths~\cite{mann3}.

\begin{figure}[H]
\centering
\includegraphics[width=7.5cm]{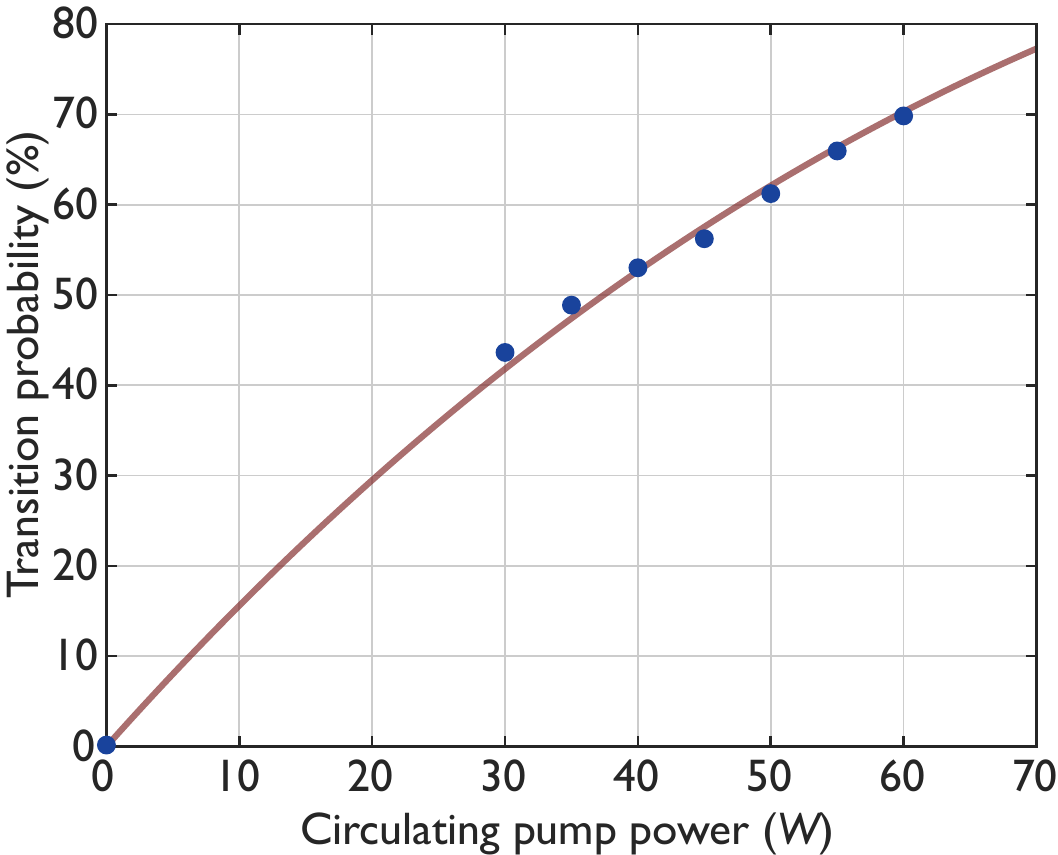}
\caption{Transition probability of the active beam splitter. The experimentally inferred transition probability was fitted with $\sin^2(\pi/2\sqrt{P_p/P_{max}})$ \cite{albota}, predicting 100\% transition probability at $P_{max}=149 \mbox{ W}$ and a 50\% transition probability at $\sim 37$W.}
\label{fig:transition_probability}
\end{figure}

To `balance' the splitting ratio of our active beam splitter, the cavity-coupled 1064 nm pump power was adjusted to a transmitted pump power of $(700\pm 6)$ mW, from which we inferred a circulating pump power of $(35\pm 0.3)$~W and a resulting transition probability of $(47.6 \pm 0.7)\%$ (Fig.~\ref{fig:transition_probability}) bounding the maximal visibility of the HOM dip to $99.4\pm 0.2\%$. The transition probability was determined from the direct measurement of the ratio of coincidence counts $p_2/p_1$ for large delay $\Delta \tau$ -- much larger than the single photon coherence length. From the total counts in each of the two distinct coincidence peaks - one indicating that both photons remained (peak 1) or both photons swapped (peak 2) we can infer $p_2/p_1 = \eta_c^2/(1-\eta_c)^2$ and obtain $\eta_c= \sqrt{p_2/p_1}/(1+\sqrt{p_2/p_1}).$

To realize the color HOM interference, the photon pair -- the 637 nm photon and 1587 nm photon -- were both aligned collinearly into the converter crystal, where they undergo SFG/DFG with the 1064~nm pump light. The input and collection modes were first optimized to maximize conversion efficiency at the given pump power. To reduce pump-induced background noise, the 637 nm light was spectrally filtered with a 20 nm bandpass filter (BPF) before collection into a single-mode fiber (SMF). The light at telecommunication wavelength was then filtered with a BPF and a monochromator, consisting of a transmission grating (TG) and a SMF, with a filter bandwidth of $\Delta\nu_{f} \approx 105\mbox{ GHz}$ (FWHM). A variable temporal delay was introduced on the telecom photon path, allowing the relative path delay -- and thus the temporal distinguishability of the photons -- to be scanned over 500 ps with step sizes down to 2.5 fs. Both red and telecom output modes were subsequently split by fiber-based balanced beam splitters and detected via two avalanche photodiodes (APDs) and two superconducting nanowire single photon detectors (SNSPDs), respectively.

Careful tuning of the wavelength of the blue pump laser and the crystal temperatures of both source and converter was required to ensure the energy difference of the photon pair matched the frequency of the pump laser of the converter while simultaneously satisfying the phase-matching condition. The measured collection efficiencies of the source were about 57\% for the red arm and 33\% for the telecom arm, and were ultimately limited by the non-degeneracy of the source alongside optical losses. The overall transmission of the converter setup was 46\%. The SNSPDs and APDs had nominal detection efficiencies of 90\% and 35\%, respectively.

Analogous to standard HOM interference, ideal two-photon interference in color demands that the (composite) wavefunction of the two photons must be described by a pure state. Here, the two single photons were created as a pair from the same SPDC source, with the resulting pure biphoton state strongly entangled in frequency. The resulting HOM interference in color is analogous to the HOM interference of a wavelength-degenerate bi-photon. To achieve strong two-photon interference, this underlying frequency entanglement shared between signal and idler must be preserved by the quantum frequency converter~\cite{mann2}.

\section{Results \& Discussion}

\begin{figure}[H]
\centering
\includegraphics[width=\textwidth]{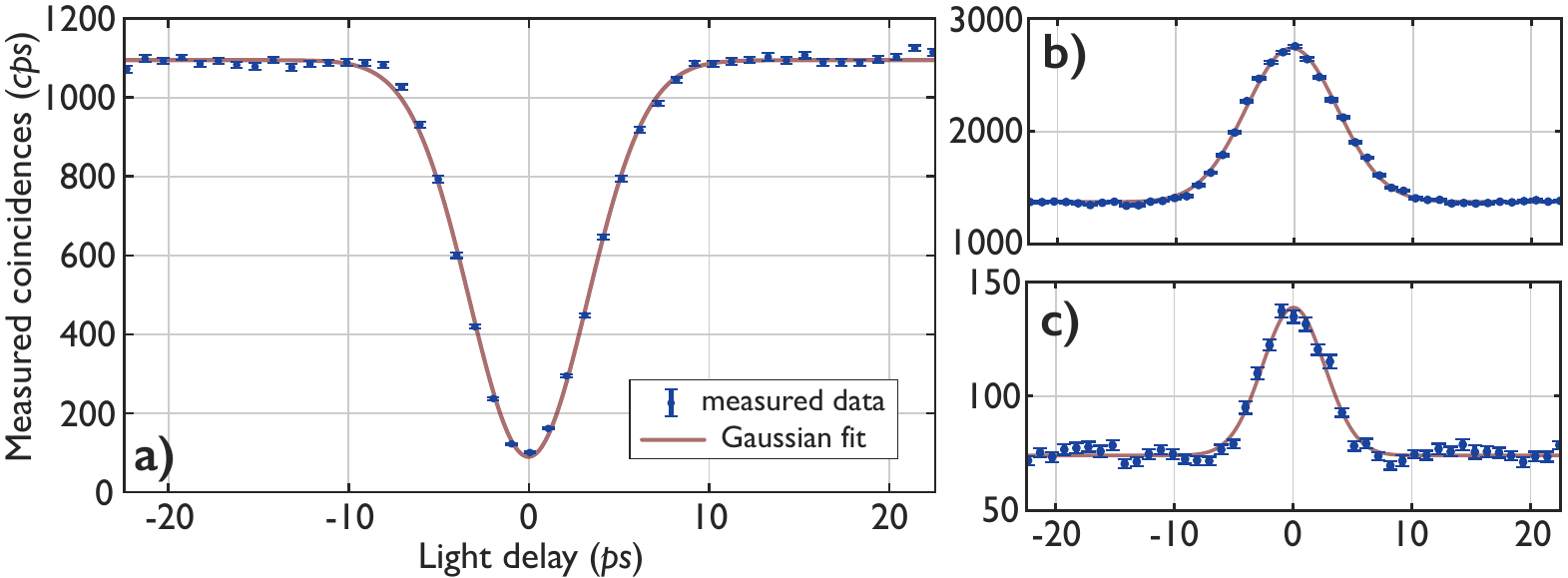}
\caption{{\bf Hong-Ou-Mandel (HOM) dip in color:} (a) measured telecom-red coincidences as a function of the optical delay in the telecom arm. The visibility of the fitted Gaussian is (91.4 $\pm$ 0.3)\% without the subtraction of accidental coincidences. The width of the fitted Gaussian is (7.78 $\pm$ 0.06) ps (FWHM). Measured (b) telecom-telecom and (c) red-red coincidences as a function of the optical delay in the telecom arm. The corresponding widths of the fitted Gaussian anti-dips are for telecom (8.99 $\pm$ 0.11) ps and for red (5.99 $\pm$ 0.34) ps.}
\label{fig:dip}
\end{figure}

Figure \ref{fig:dip} shows the measured two-color HOM dip and single-color anti-dips. In a single measurement, the time-tags from the four detectors were recorded as a function of the relative temporal delay introduced in the telecom arm before the detector. The HOM dip in Figure \ref{fig:dip}a) results from correlating the cumulated time-tags from the two SNSPDs with the cumulated time-tags from the two APDs, achieving a raw visibility of (91.4 $\pm$ 0.3)\%. This is well above the classical limit of 50\% visibility~\cite{ou_book}. With the subtraction of accidental coincidences, the measured visibility improves to (92.8 $\pm$ 0.3)\%. We assume the bandwidth of the converter to be the primary limitation to the measured HOM visibility. It results in a small frequency dependence in the swap and no-swap terms of the active beam splitter. As not all frequency components of a photon experience the ideal 50:50 splitting ratio of the beam splitter, the resulting imbalance introduces some path distinguishability. This effect can also be viewed as a spectral deformation of the photons, with the photons undergoing some small spectral reshaping, depending on whether they are converted or unconverted. The measured temporal width of the dip is 7.78 $\pm$ 0.06 ps (FWHM), with a predominantly Gaussian shape, owing to spectral filtering in the telecom arm and the limited converter bandwidth. Both filters suppress the sidelobes of the sinc function describing the spectral distribution of the biphoton, broadening the dip in time. The anti-dips in Figure~\ref{fig:dip} b) and c) result from individually correlating the time-tags from the b) SNSPDs c) APDs. The corresponding widths (and visibilities) of the fitted Gaussians to the anti-dips are (8.99 $\pm$ 0.11) ps ($\mbox{Vis}=(99.8\pm1.2)\%$) for telecom-telecom coincidences and (5.99 $\pm$ 0.34) ps ($\mbox{Vis}= (81.1\pm 4.0)\%$) for the red-red coincidences. The different temporal widths and higher visibility for the telecom-telecom anti-dip arise primarily from stronger spectral filtering in the telecom arm.

This experiment marks the first realization of color HOM interference with vastly different energies for the participating single photons, with an octave-spanning energy difference of 282 THz, exceeding previous experiments by more than two orders of magnitude. Resolving the beat note of two photons~\cite{legero} of such fundamentally different color with a passive beam splitter would require a detection system with a temporal resolution below $\Delta t_{beat}= 1/\delta \nu \approx 3.5 \mbox{ fs}$. Such a timing resolution is orders of magnitude away from currently available technology. Moreover, it constitutes the first realization of color HOM interference where the energy difference of the participating single photons is comparable to those typical of hybrid, fiber-based quantum networks, where single-emitters with excitation energies corresponding to visible to near-infrared wavelengths are interconnected with telecommunications wavelengths.

Beyond extending our understanding of what underlies the phenomena of quantum interference, the possibility to interfere photons of such different color also has implications for quantum information processing~\cite{review_gaeta,lukens2016,fabre2022}. Traditionally, when building a quantum network, the challenge of interfacing photonic qubits -- usually at telecommunication wavelengths -- with heterogeneous quantum systems would see frequency conversion as an intermediate step to ensure wavelength compatibility. Instead, we suggest the use of the converter as the beam splitter itself. This work also complements growing frameworks in photonic quantum information processing, where encodings, evolution and detection exploit the spectral~\cite{review_gaeta} and temporal-spectral~\cite{BrechtPRX} degrees-of-freedom. 

\section{Conclusion}
In conclusion, we have experimentally demonstrated HOM interference with two single photons of vastly different color. Using a quantum frequency converter as an active beam splitter, we coupled two single photons - at 637 nm and 1587 nm - via a strong 1064 nm pump field. We observed a strong suppression in coincidence counts between the two color modes, achieving a HOM dip visibility of (91.4 $\pm$ 0.3)\% -- well beyond the classical limit of 50\%. This work experimentally highlights the role of path indistinguishability in quantum interference and adds a new element to the toolbox for quantum technology applications. 


\begin{backmatter}

\bmsection{Funding} Funded by the BMBF, Germany within the project QR.N (16KIS2185).


\bmsection{Disclosures} The authors declare no conflicts of interest.

\bmsection{Data availability} Data underlying the results presented in this paper are not publicly available at this time but may be obtained from the authors upon reasonable request.

\bmsection{Supplemental document} See Supplement 1 for supporting content.

\end{backmatter}

\vspace{1cm}
\setcounter{section}{0}

\title{Supplemental document}
\author{} 

\textbf{Abstract:} In this supplemental document, we recap quantum frequency conversion \cite{kumar} as a beam splitter operation, the quantum mechanical treatment of passive and active lossless beam splitters \cite{campos,raymer,review_gaeta}, and the Hong-Ou-Mandel effect. While for a passive beam splitter all four modes have the same frequencies, an active beam splitter allows for different frequencies of the participating modes.\\ 
Already in the treatment by Campos et al. \cite{campos}, only linear input-output relations and bosonic commutation rules for the input and output modes are assumed. These assumptions are sufficient to fix the relative phase between the modes to $\pi$ and to show unitarity and particle number conservation for the beam splitter operation. Energy conservation is just given as a consequence of particle number conservation for the case of all participating modes having the same frequencies. In the publication by Raymer et al. \cite{raymer}, the concept of the active beam splitter is introduced and various physical implementations are discussed.

\section{Quantum frequency conversion}
The single-mode interaction Hamiltonian for three-wave-mixing can be written as \cite{kumar}
\begin{gather}
    \hat{H}_I= i \hbar \chi'(\hat{a}_1 \hat{a}_p \hat{a}_2^\dagger - h.c.)
\end{gather}
Assuming a classical pump mode we have
\begin{gather}
    \hat{H}_I= i \hbar \chi(\hat{a}_1  \hat{a}_2^\dagger - h.c.)
\end{gather}
with $\chi=\chi'E_p$. The Heisenberg equations of motion in the interaction picture lead to the solution \cite{review_gaeta}
\begin{align}
    \hat{b}_1 &:= \hat{a}_1(t) = \cos(\chi t)\,\hat{a}_1(0) + i \sin(\chi t)\,\hat{a}_2(0), \\
    \hat{b}_2 &:=\hat{a}_2(t) = i \sin(\chi t)\,\hat{a}_1(0) + \cos(\chi t)\,\hat{a}_2(0).
\end{align}
Here we can define a transmittance $t := \cos(\chi t)$ and a reflectance $r := \sin(\chi t).$

\section{Passive and active quantum mechanical beam splitter}
We treat a lossless beam splitter with two input modes and two output modes. The modes could be, for example, spatial or spectral modes - the theoretical treatment is fully equivalent. We start by assuming a linear input-output relationship for the annihilation operators \cite{campos}
\begin{gather}
\begin{pmatrix}
\hat{b}_1\\
\hat{b}_2
\end{pmatrix}
=
\begin{pmatrix}
B_{11} & B_{12} \\
B_{21} & B_{22} 
\end{pmatrix}
\begin{pmatrix}
\hat{a}_1\\
\hat{a}_2
\end{pmatrix}
.
\end{gather}
The four elements of the beam splitter matrix are in general complex numbers and can be written as
\begin{gather}
    B_{ij}=|B_{ij}|e^{i\phi_{ij}} \mbox{\hspace{0.5cm}} i,j=1,2. \label{eq:beam_splitter_polar}
\end{gather}
We calculate
\begin{gather}
   [\hat{b}_1,\hat{b}_1^{\dagger}]=1 \hspace{0.5cm}\Rightarrow |B_{11}|^2+|B_{12}|^2=1,\label{eq:beam_splitter_a}\\
   [\hat{b}_2,\hat{b}_2^{\dagger}]=1 \hspace{0.5cm}\Rightarrow|B_{21}|^2+|B_{22}|^2=1,\label{eq:beam_splitter_b}\\
   [\hat{b}_1,\hat{b}_2^{\dagger}]=0 \hspace{0.5cm}\Rightarrow B_{11}B_{21}^*+B_{12}B_{22}^*=0. \label{eq:beam_splitter_c}
\end{gather}
We insert equation \ref{eq:beam_splitter_polar} into equation \ref{eq:beam_splitter_c} and find 
\begin{gather}
    |B_{11}||B_{21}|=|B_{12}||B_{22}|, \label{eq:beam_splitter_magnitudes}\\
   \phi_{11}-\phi_{21}=\phi_{12}-\phi_{22} \pm \pi \label{eq:beam_splitter_phases}.
\end{gather}
The system of equations \ref{eq:beam_splitter_magnitudes}, \ref{eq:beam_splitter_a} and \ref{eq:beam_splitter_b} enforce
\begin{gather}
    |B_{11}|=|B_{22}| =: t, \hspace{1cm}|B_{12}|=|B_{21}| =: r.
\end{gather}
Here t is the transmittance and r is the reflectance of the beam splitter - which are both real numbers. Inserting the above finding into equation \ref{eq:beam_splitter_a} (or \ref{eq:beam_splitter_b}) we arrive at
\begin{gather}
    t^2+r^2=1.
\end{gather}
With a convenient choice of phases ($\phi_{11}=\phi_{22}=0$, $\phi_{21}=\phi_{12}=\pi/2$)  - satisfying equation \ref{eq:beam_splitter_phases}  - we have
\begin{gather}
    \hat{B}=
\begin{pmatrix}
t & ir \\
ir & t 
\end{pmatrix}
.
\end{gather}
It is easy to verify that this matrix is a unitary transformation $\hat{B}^\dagger \hat{B} = \mathbb{1}$. Particle number conservation holds in the form of the equation $\hat{N}_1+\hat{N}_2=\hat{n}_1+\hat{n}_2$, with $\hat{N}_j = \hat{b}_j^{\dagger}\hat{b}_j$ and $\hat{n}_j = \hat{a}_j^{\dagger}\hat{a}_j$. If the modes have the same frequency $\omega$, energy conservation follows from the particle number conservation equation above - via multiplying it by $\hbar \omega$. In that sense the passive beam splitter is a special case of the active beam splitter. Often one does not just want to map the input on the output operators - but map the output on the input operators as
\begin{gather}
\begin{pmatrix}
\hat{a}_1\\
\hat{a}_2
\end{pmatrix}
=
\hat{B}^{-1}
\begin{pmatrix}
\hat{b}_1\\
\hat{b}_2
\end{pmatrix}
.
\end{gather}
Since $\hat{B}$ is a unitary transformation this inverse operation is just given by the Hermitian conjugate $\hat{B}^{-1}=\hat{B}^{\dagger}$. Further the matrix $\hat{B}$ is symmetric and thus $\hat{B}^{\dagger}=\hat{B}^{*}$. We thus find $\hat{B}^{-1}=\hat{B}^{*}$, and hence
\begin{gather}
    \hat{a}_1 = t\hat{b}_1-ir\hat{b}_2,\\
    \hat{a}_2 = -ir\hat{b}_1+t\hat{b}_2.
\label{eq:BS_annihilation}
\end{gather}
We apply the dagger operation to the above equations and set $t=r=\frac{1}{\sqrt{2}}$ for a balanced beam splitter. Thus, the output-input relations in terms of the creation operators are given as \cite{raymer}
\begin{gather}
    \hat{a}^\dagger_1 = \frac{1}{\sqrt{2}}\left(\hat{b}^\dagger_1+i\hat{b}^\dagger_2\right),\label{eq:BS_creation1} \\
    \hat{a}^\dagger_2 = \frac{1}{\sqrt{2}}\left(i\hat{b}^\dagger_1+\hat{b}^\dagger_2\right). \label{eq:BS_creation2}
\end{gather}

\section{Hong-Ou-Mandel effect for passive and active beam splitters}

We assume the input state  \cite{raymer} 
\begin{gather}
\ket{\psi_{\text{in}}} = a_1^\dagger a_2^\dagger \ket{0}.
\end{gather}
Applying the beam splitter operations \ref{eq:BS_creation1} and \ref{eq:BS_creation2} gives
\begin{gather}
\ket{\psi_{\text{out}}}
= \frac{1}{2}\!\left(\hat{b}_1^\dagger + i\,\hat{b}_2^\dagger\right)\!\left(i\,\hat{b}_1^\dagger + \hat{b}_2^\dagger\right)\ket{0}, \\
= \frac{1}{2}\!\left(
i\,\hat{b}_1^{\dagger 2}
+ \hat{b}_1^\dagger \hat{b}_2^\dagger
- \hat{b}_2^\dagger \hat{b}_1^\dagger
+ i\,\hat{b}_2^{\dagger 2}
\right)\ket{0}
= \frac{i}{2}\left(\hat{b}_1^{\dagger 2}+\hat{b}_2^{\dagger 2}\right)\ket{0}.
\end{gather}



\end{document}